\RequirePackage{fix-cm}
\documentclass[twocolumn,epjc3]{svjour3}  
\smartqed  
\usepackage{color}
\usepackage{amssymb}
\usepackage{amsmath}
\RequirePackage{graphicx}

\journalname{Eur. Phys. J. C}
\begin{document}

\title{Characteristics of extensive air showers around the energy threshold for ground-particle-based $\gamma$-ray observatories}


\author{Harm Schoorlemmer\thanksref{addr1,e1}
        \and
        Jim Hinton\thanksref{addr1} 
        \and
        Rub\'en L\'opez-Coto\thanksref{addr1, addr2} 
}

\thankstext{e1}{e-mail: harmscho@mpi-hd.mpg.de}


\institute{Max-Planck-Institut f\"ur Kernphysik, Saupfercheckweg 1, 69117 Heidelberg, Germany \label{addr1}
           \and
           now at Universit\`{a} di Padova and INFN, I-35131, Padova, Italy \label{addr2}
}

\date{Received: date / Accepted: date}

\maketitle

\begin{abstract}
  Very high energy $\gamma$-ray astronomy based on the measurement of air shower particles at ground-level has only recently been established as a viable approach, complementing the well established air Cherenkov technique. This approach requires high (mountain) altitudes and very high surface coverage particle detectors. While in general the properties of air showers are well established for many decades, the extreme situation of ground-level detection of very small showers from low energy primaries has not yet been well characterised for the purposes of $\gamma$-ray astronomy. Here we attempt such a characterisation, with the aim of supporting the optimisation of next-generation $\gamma$-ray observatories based on this technique. We address all of the key ground level observables and provide parameterisations for use in detector optimisation for shower energies around 1~TeV. We emphasise two primary aspects: the need for large area detectors to effectively measure low-energy showers, and the importance of muon identification for the purpose of background rejection.

\end{abstract}

\section{Introduction} 

In the past two decades VHE $\gamma$-ray astronomy has become established as a rich astronomical discipline, based primarily on observations with Imaging Atmospheric Cherenkov Telescope (IACT) arrays. However, given the limitations of IACTs in terms of Field-of-View (FoV, typically a few degrees diameter) and duty cycle (limited to typically $10-15\%$), new approaches are being developed to provide very large field of view and continuous observations. The direct detection of extensive air shower (EAS) particles at ground level is the most obvious complementary approach.
Air Shower Particle Detectors (ASPDs), can observe a large fraction of the overhead sky at any given time and, with their close to 100\% up-time, can survey roughly 2/3 of the sky on a daily basis. However, since they measure only the ground-level properties of air showers, the accuracy with which they reconstruct the properties of the primary $\gamma$-ray is typically significantly worse than is the case for IACTs. Typical ASPD (IACT) performance values around 1~TeV are angular resolution 0.4$^{\circ}$ (CF 0.06$^{\circ}$), hadron rejection efficiency 99\% (CF 99.9\%) and energy resolution of 50\% (CF 10\%) --- see~\cite{HAWC_CRAB,HESS_Performance,MAGIC_Performance,impact,HAWC_Energy}.

During their propagation through the atmosphere, the number of particles
in the air shower grows due to production of secondary particles until
an atmospheric depth $X_\text{max}$. After this maximum, the energy content of the shower declines as ionisation
losses dominate over new particle production. ASPDs are typically
located at high altitude sites to be as close to $X_\text{max}$ as
possible and maximise the particle count per shower, improving the
detection efficiency and accuracy. Ground-based $\gamma$-ray astronomy
relies on discrimination of $\gamma$-rays from the dominant background of
cosmic ray protons and nuclei on the basis of air shower
characteristics.
The most obvious differentiating characteristic of hadronic versus purely electromagnetic (EM) cascades is the presence of pions, and subsequent production of muons in pion decay as well as EM sub-showers. For a recent detailed review of the properties of air showers we refer the reader to \cite{engel_air_showers}.

Several instrumentation approaches have been used for ASPDs at different observatories around the world: the Tibet AS-$\gamma$ experiment~\cite{Tibet_Survey} used scintillator-based detectors, ARGO-YBJ is based on Resistive Plate Chambers (RPCs)~\cite{ARGO_Performance_2006}, but up to now, the most successful technique in terms of sensitivity, angular and energy resolution is the water Cherenkov technique introduced in MILAGRO~\cite{Milagro_Crab} and currently implemented in the High Altitude Water Cherenkov (HAWC) $\gamma$-ray observatory~\cite{HAWC_Performance_2013}. HAWC has reached the critical sensitivity required to detect $\gamma$-ray sources in significant numbers~\cite{HAWC_survey}.
There is now a significant push towards an ASPD in the Southern Hemisphere, with a wide range of technologies being explored
\cite{LATTES_ICRC,ALTO_ICRC,ALPACA_ICRC,ARGO_ICRC,HARM_ICRC,SCIENCECASE_ICRC}.
Here we study the properties of $\gamma$-ray and background showers at around the threshold where the ASPD approach becomes possible for a high altitude detector, focussing on aspects that influence the design and utilisation of such a detector. 


\section{ Simulations \& Atmospheric Propagation} 
\label{sec:shower_physics}

$\gamma$-ray and proton induced air showers are simulated using the CORSIKA (COsmic Ray SImulations for KAscade) 7.4005 package\cite{corsika}. Within the CORSIKA package, the hadronic interaction event generator FLUKA is used, in combination with the QGSJet-II model for high-energy interactions, with electromagnetic interactions based on
EGS4.  A set of fixed energies in the range between 50~GeV and 5~TeV and several fixed zenith angles in the range between 0$^\circ$ and 60$^\circ$ are simulated.
For each energy and zenith angle a range of detector elevations from 3.5~km to 6.0~km above sea level are considered, covering the plausible range for a future ASPD observatory. For the charged background of cosmic nuclei we consider only protons, as they likely provide the dominant source of background events in ASPD $\gamma$-ray observatories in this energy range.

The parameter that drives the number of interactions and energy losses in the shower development is the amount of material that the particles in the EAS traverse during their propagation through the atmosphere, the {\it{slant depth}}, given by the line integral over the atmospheric density $\rho_{\text{atm}}$  from a location of altitude $z$ above sea level in the direction of shower origin:
$X(z,\theta) = \int_\infty^{z}\rho_{\mathrm{atm}}(z',\theta)\cos\theta dz'$,
where $\theta$ is the zenith angle.
Figure \ref{fig:slant_depth} in Appendix~A provides a reference for the relationship of slant depth to altitude ($z$) and zenith angle $\theta$ for the {\it US standard atmosphere} density profile, for use in interpreting the results given gere in terms of slant depth.
At all feasible altitudes for an observatory in this energy range, shower maximum typically occurs well before the shower reaches the ground. The dependency of the average value of $X_{\text{max}}$ on energy, usually called the elongation rate, can be parameterised as $\langle X_{\text{max}} \rangle = a + b \log_{10}(E/\text{GeV})$ \cite{Heitler_model,Rossi_model},
where $E$ is the energy of the primary particle. For simulated $\gamma$-ray showers we find $a = 98$~g~cm${^{-2}}$ and $b=83$~g~cm${^{-2}}$ and for proton primaries $a = 111$~g~cm${^{-2}}$ and $b =74$~g~cm${^{-2}}$. This means that for an observatory at an altitude of 5~km a 100~GeV (1~TeV) $\gamma$-ray induced shower from zenith is on average 7.7 (5.5) radiation lengths beyond the atmospheric depth at which it reached $X_{\text{max}}$, implying very few particles reaching ground-level and large fluctuations.   

\section{Ground-level Particles}  

\begin{figure}[!ht]
\begin{center}
\includegraphics[width=0.5\textwidth]{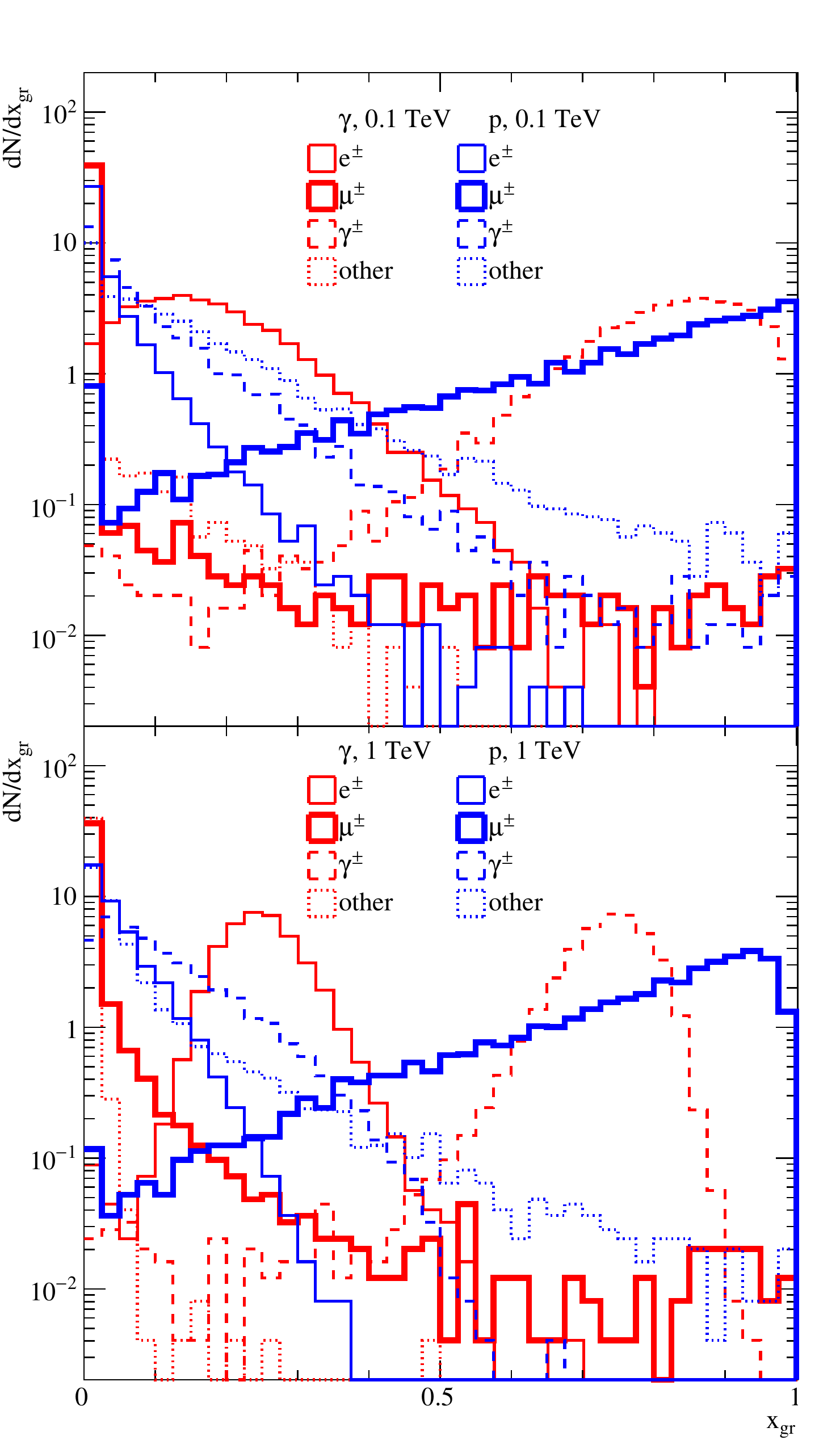}
\caption{Distribution of the fraction $x_{\text{gr}} = E_i/E_{\text{gr}}$ of the total energy that reaches ground $E_{\text{gr}}$ per particle type $i$. $\gamma$-ray and proton induced vertical EASs with 100~GeV (top panel) and 1~TeV energies (bottom panel) are shown for an altitude of 5~km above sea level.}
\label{fig:frac_part_en}
\end{center}
\end{figure}

The vast majority of the energy in EASs is carried to the ground by photons, electrons and positrons (hereafter simply electrons), and muons ($\mu^{+}$ and $\mu^{-}$). Figure~\ref{fig:frac_part_en} shows for each of these particle types the distribution of the fraction of the total energy arriving at the ground ($x_{\text{gr}} = E_{i}/E_{\text{gr}}$) for 100~GeV and 1~TeV proton and $\gamma$-ray primaries. For $\gamma$-ray showers the well-known dominance of photons is apparent, as well as the occasional production of (and even very occasionally the energetic dominance at the ground of) muons. For proton initiated showers the majority of the arriving energy is typically in the form of muons. For both primary particle types huge fluctuations are evident at these energies, as expected due to the small number of particles reaching the ground. We note that the category ``other'' is dominated by protons and neutrons, which can carry a significant fraction of the energy in rare cases.

\begin{figure}[!ht]
\begin{center}
\includegraphics[width=0.5\textwidth]{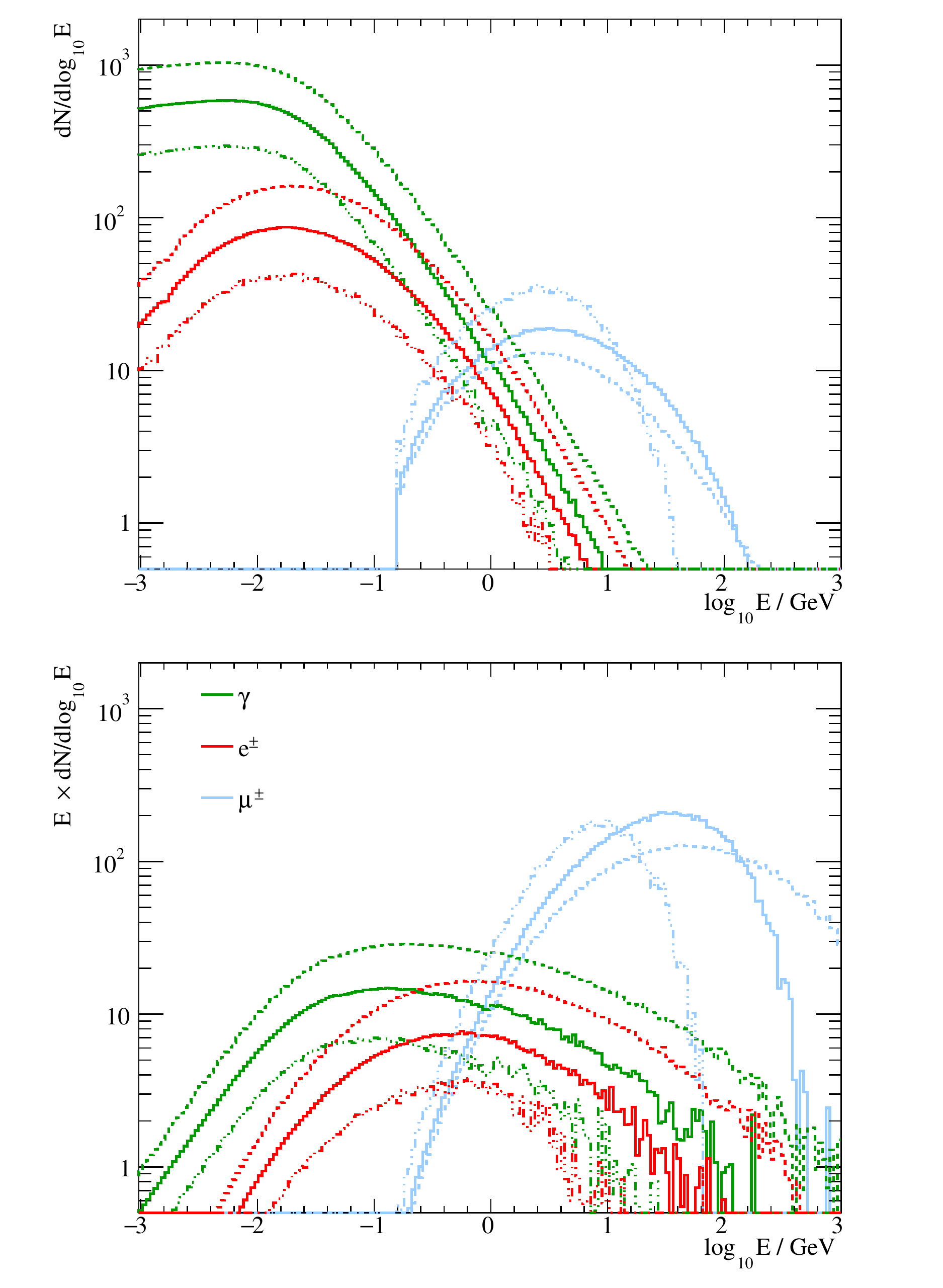}
\caption{Number density (top) and energy density distributions (bottom) of ground-level particles for showers initiated by primary protons of three energies, solid lines correspond to 1~TeV primary energy, dashed lines to 10~TeV, and dashed-dotted lines to 100~GeV. The distributions of the 10~TeV case have been scaled down by a factor of ten, while the 100~GeV curves have been multiplied by a factor of ten.}
\label{fig:part_spec}
\end{center}
\end{figure}

The distributions of individual particle energies at ground-level are very different for muons with respect to electrons and photons, as illustrated in Figure~\ref{fig:part_spec}.  The peak of the energy distribution in terms of number per log energy interval($dN/d\log E$) is around $\sim$6~MeV for photons, $\sim$20~MeV for electrons and 2--3~GeV for muons. In terms of total energy per log interval ($EdN/d\log E$) the peak lies at $\sim$150~MeV for photons, $\sim$600~MeV for electrons and 30--40~GeV for muons.

\begin{figure}[!ht]
\begin{center}
\includegraphics[width=0.5\textwidth] {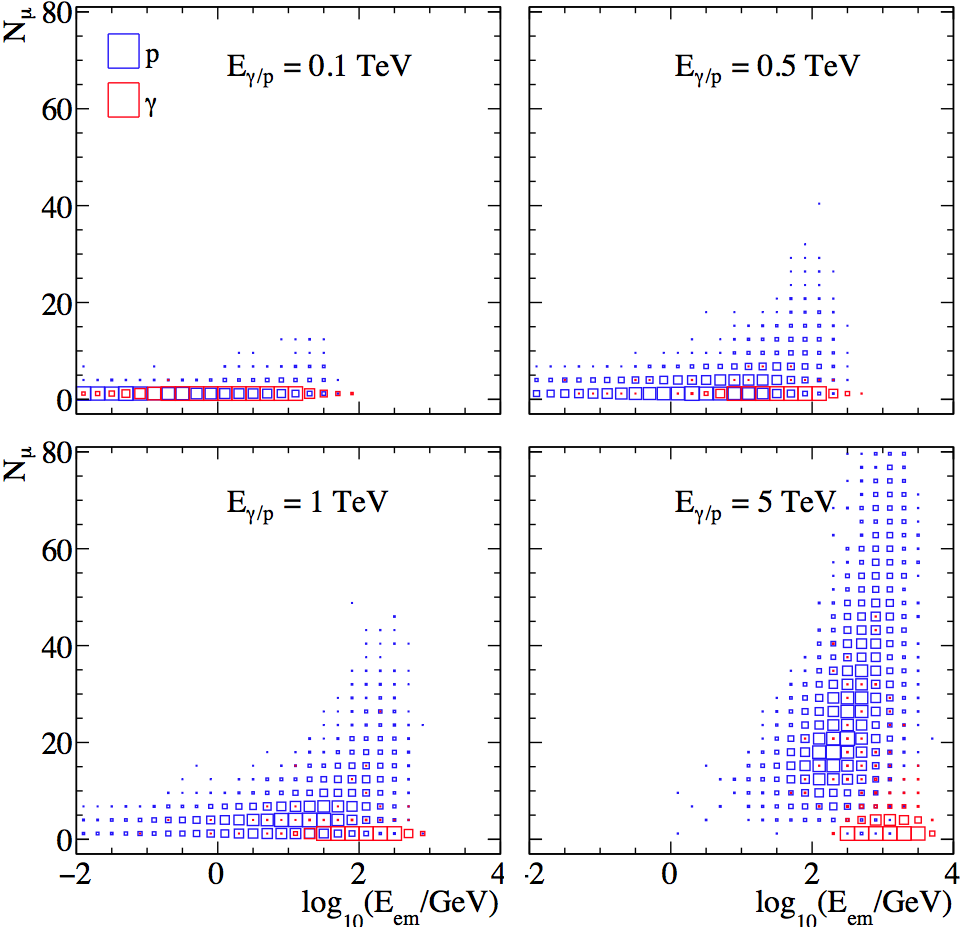}
\caption{Distributions of the number of muons within 100~m of the impact point as a function of ground-level electromagnetic energy for vertical showers observed at 5~km altitude.
  Proton (blue) and $\gamma$-ray (red) initiated showers of different energies are shown, with the area of the squares scaling linearly with density.} 
\label{fig:fNmu_Eem_distr}
\end{center}
\end{figure}

Many ASPDs make use of calorimetric detectors that are sensitive to the total ground-level electromagnetic energy ($E_{\text{em}}$), rather than the number of electrons. Muons, in contrast, rarely loose a significant amount of their energy in typical detectors and hence their number ($N_{\mu}$) is the relevant quantity. 
Figure \ref{fig:fNmu_Eem_distr} shows the distributions of muon number $N_{\mu}$ versus electromagnetic energy $E_{\text{em}}$ for proton and $\gamma$-ray initiated showers of several energies.
A large difference in $N_\mu$ at fixed $E_{\text{em}}$ between $\gamma$-ray and proton induced EASs is apparent, with useful separation power appearing at energies above about 1~TeV provided both of these quantities can be adequately measured.

\section{Parameterisation of Electromagnetic Energy at Ground}  

\begin{figure}[!ht]
\begin{center}
\includegraphics[width=0.5\textwidth] {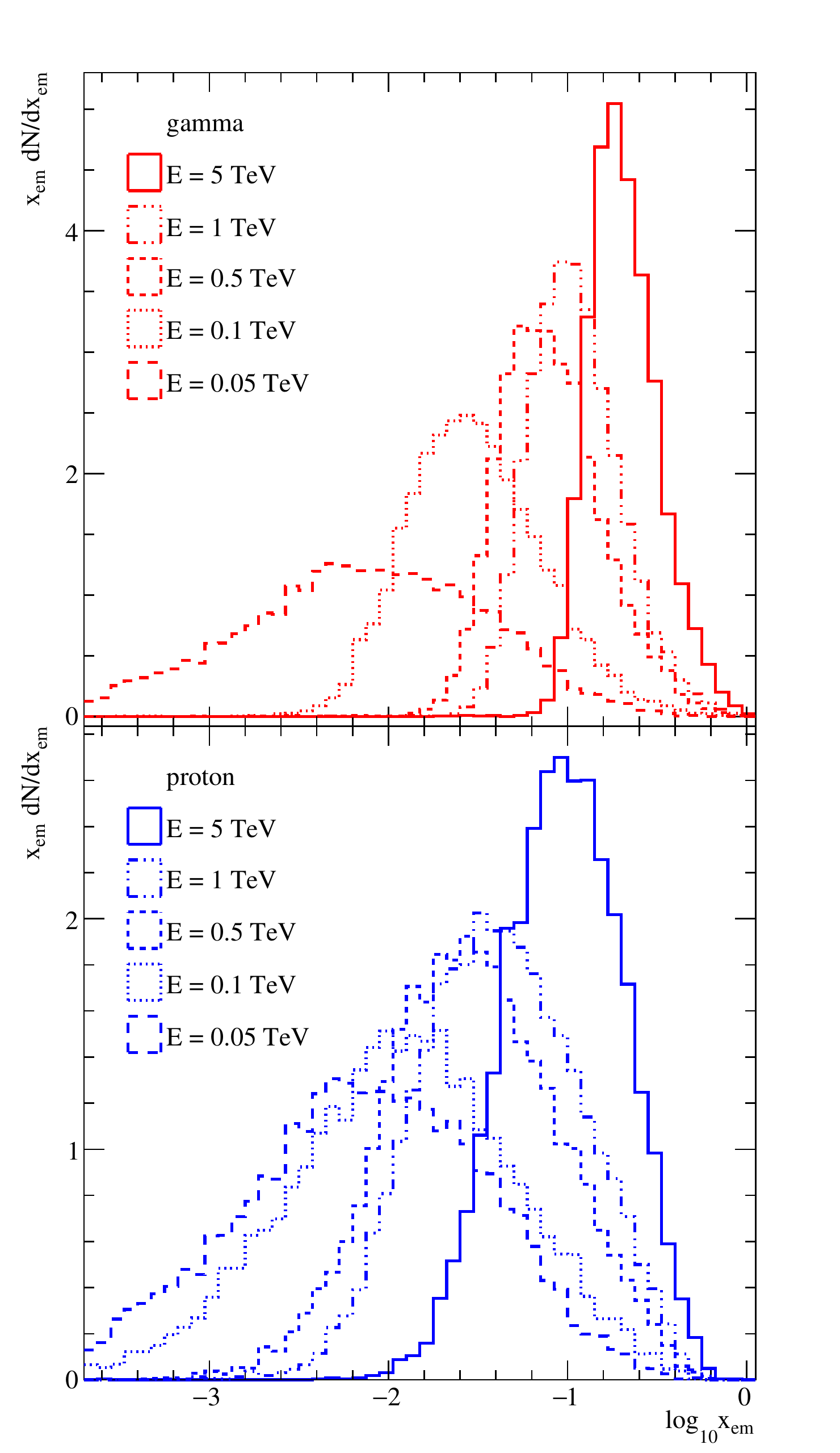}

\caption{Distributions of the electromagnetic energy that reaches the ground at 5~km altitude for vertical showers, expressed as a fraction of the primary particle energy $x_{\text{em}} = E_{\text{em}} / E$. $\gamma$-ray initiated showers are shown in the top panel and protons in the bottom panel.}
\label{fig:EmDist_50}
\end{center}
\end{figure}

Figure~\ref{fig:EmDist_50} shows the distribution of the fraction of the primary energy which reaches the ground in the electromagnetic part of EASs ($x_{\text{em}}$) for different energies and primaries.
It is clear that for proton-initiated showers not only does less electromagnetic energy reach the ground, but that fluctuations from shower to shower are a lot larger than for $\gamma$-ray showers. This is especially the case for the lowest energy showers considered, where just a few muons may carry the majority of the total energy at ground. In addition, we observe that the quantity $\log_{10} (x_{\rm{em}})$ follows an approximately Gaussian distribution. Therefore, the distributions in $\log_{10} (x_{\rm{em}})$ can be parametrised reasonably well by only the root mean square (rms) $\sigma (\log_{10} (x_{\rm{em}}))$ and the mean $\langle \log_{10} (x_{\rm{em}})\rangle$ of the distributions. 

\begin{figure}[!ht]
\begin{center}
\includegraphics[width=0.47\textwidth] {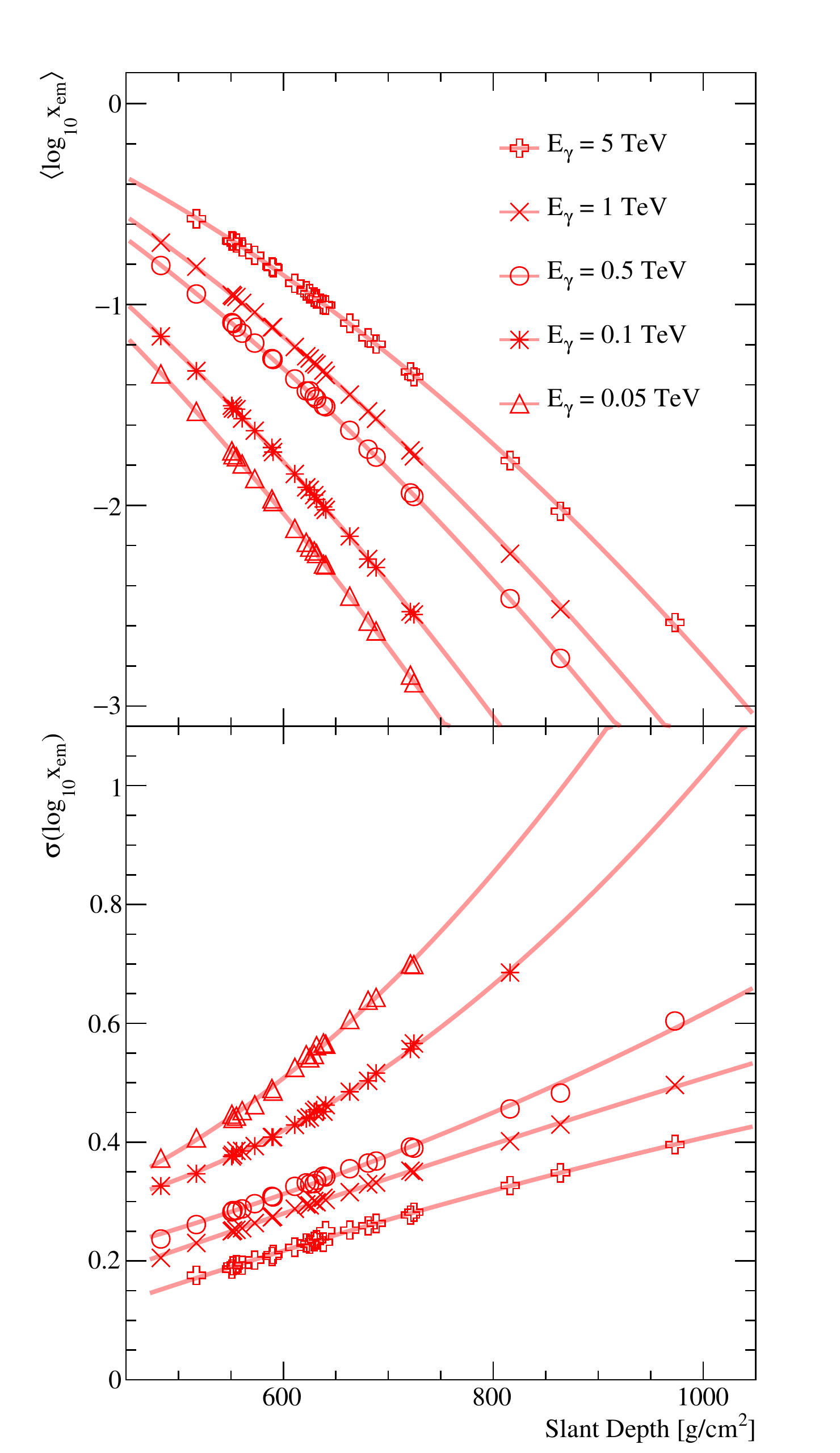}
\caption{Dependence of fractional electromagnetic energy arriving at ground $x_{\rm{em}} = E_{\text{em}} / E_{\rm \gamma}$ on slant depth, for primary $\gamma$-rays of different energies. The top panel shows the dependence of the mean of the $\log_{10} (x_{\rm{em}})$ distribution, while the bottom panel shows the behaviour of the rms ($\sigma$) of the same distribution.}
\label{fig:meanSlantDepthGamma}
\end{center}
\end{figure}

\begin{figure}[!ht]
\begin{center}
\includegraphics[width=0.47\textwidth] {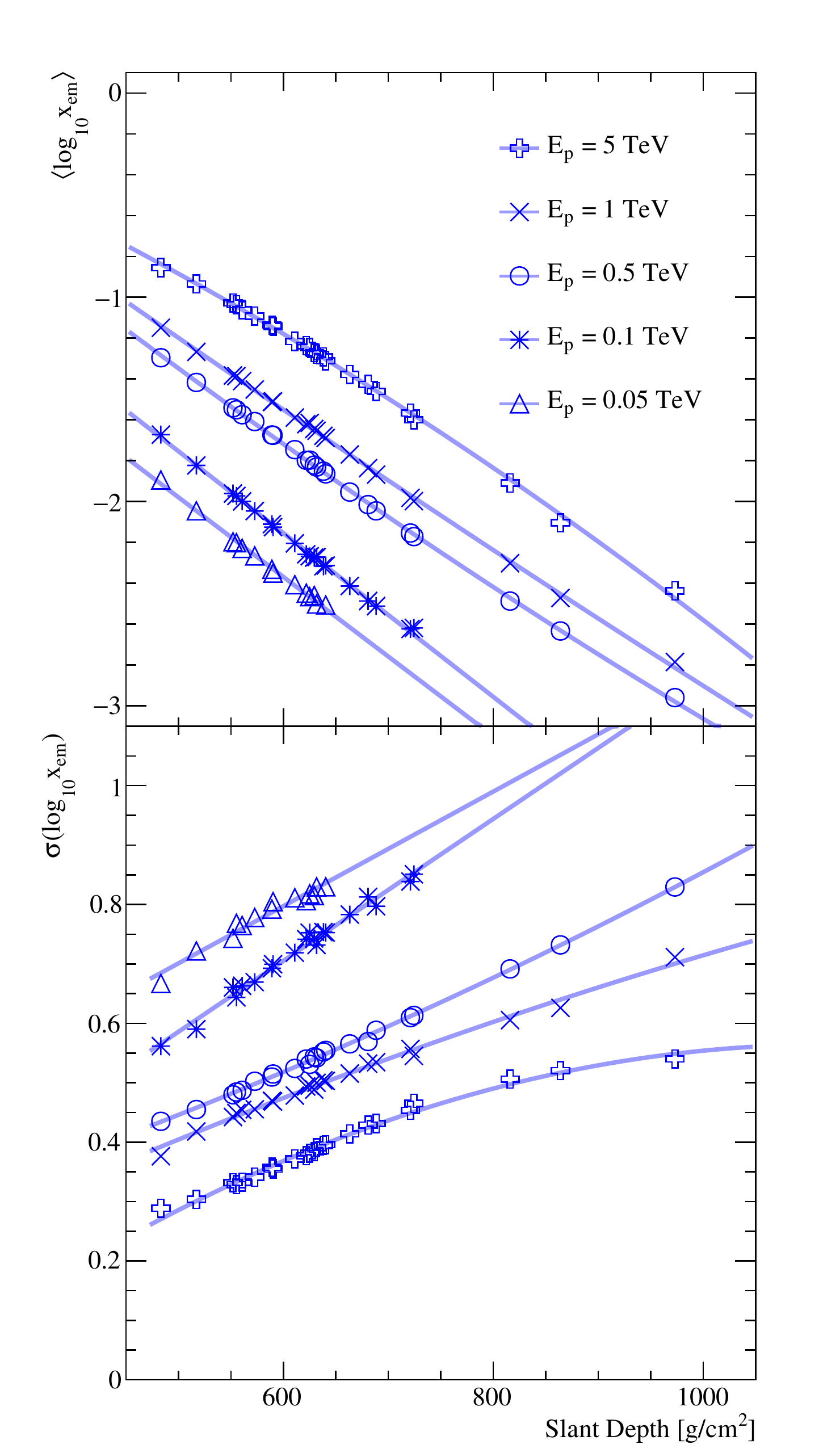}
\caption{
Dependence of fractional electromagnetic energy arriving at ground $x_{\rm{em}} = E_{\text{em}} / E_{\rm p}$ on slant depth, for primary protons of different energies. The top panel shows the dependence of the mean of the $\log_{10} (x_{\rm{em}})$ distribution, while the bottom panel shows the behaviour of the rms ($\sigma$) of the same distribution.}
\label{fig:meanSlantDepthProton}
\end{center}
\end{figure}

The resulting characterisation of the distributions of shower $x_{\text{em}}$ as a function of the slant depth is shown in Figure~\ref{fig:meanSlantDepthGamma} for $\gamma$-ray, and in Figure~\ref{fig:meanSlantDepthProton} for proton, initiated showers. As expected, the mean $E_{\rm{em}}$ decreases rapidly with increasing slant depth. For the lowest energy showers this results in a fraction of simulated showers with no particles arriving at the ground. If this fraction is more than 10\% of the total sample, we omit the sample completely from Figures \ref{fig:meanSlantDepthGamma} and \ref{fig:meanSlantDepthProton}. The dependencies of the mean and the width of the distributions on slant depth are reasonably well described by first or second order polynomials as illustrated by the fit functions represented by the lines in Figures \ref{fig:meanSlantDepthGamma} and \ref{fig:meanSlantDepthProton}. The fit parameters are given in Appendix~B. 

\begin{figure}[!ht]
\begin{center}
\includegraphics[width=0.45\textwidth] {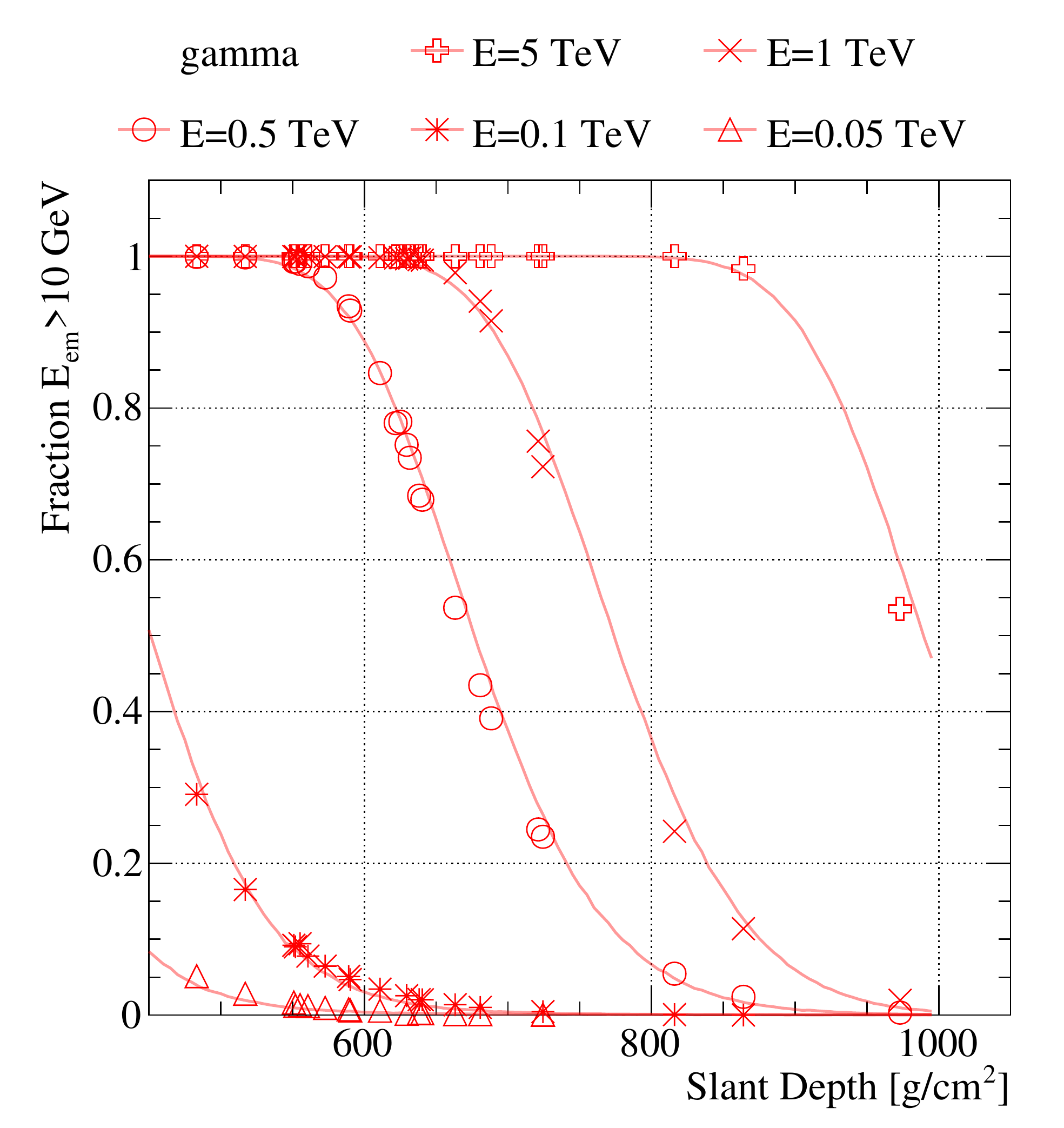}
\includegraphics[width=0.45\textwidth] {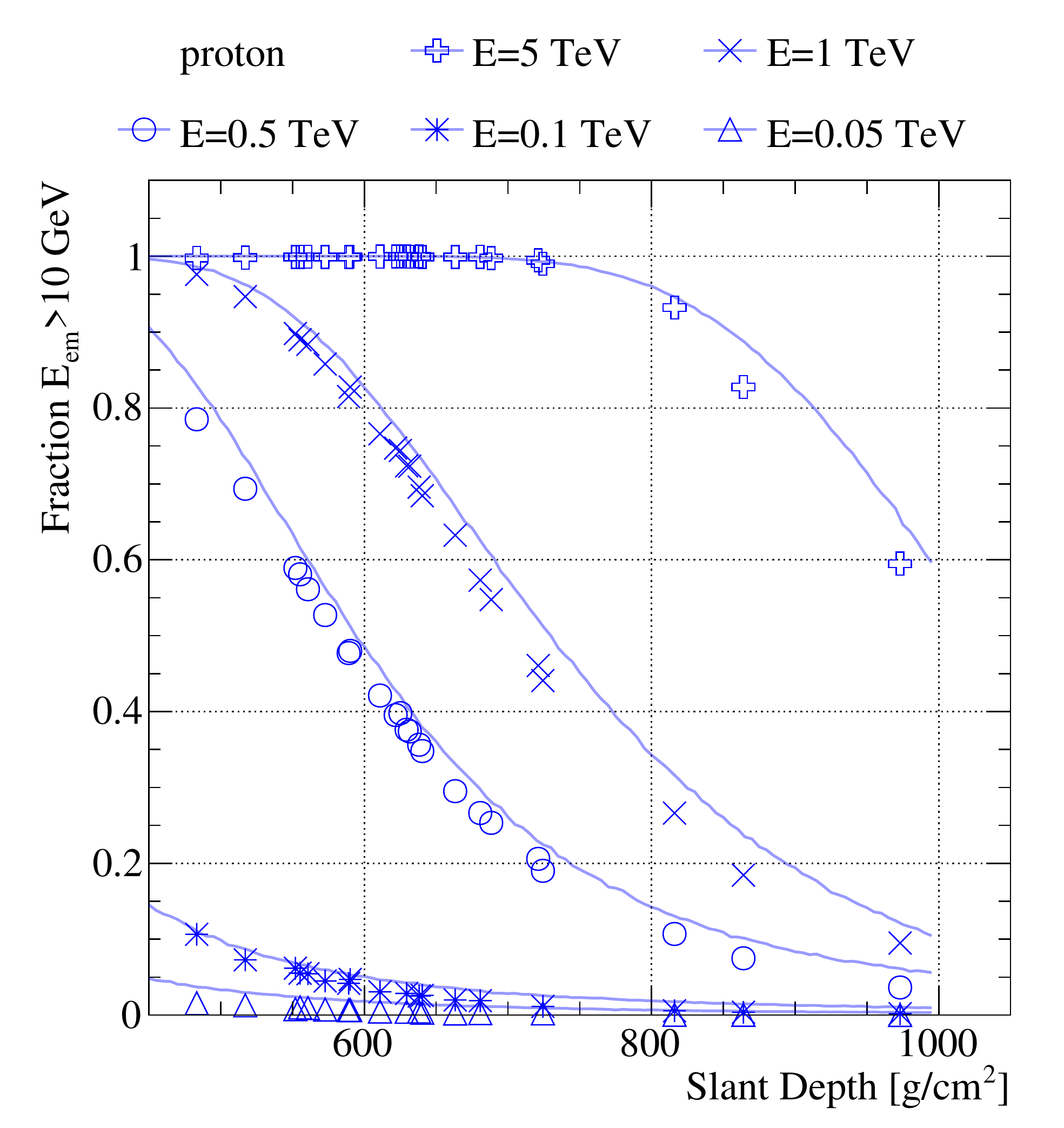}
\caption{The fraction of showers that have $E_{\text{em}}>10$~GeV as a function of slant depth. Top: $\gamma$-ray induced showers. Bottom: proton induced showers. The markers are obtained from the full simulation set, while the lines are obtained using the parameterization based on the fitted behaviour from Figure \ref{fig:meanSlantDepthGamma} and \ref{fig:meanSlantDepthProton}.}
\label{fig:Thres10GeV}
\end{center}
\end{figure}

Well above detection threshold effects, the collection area of an ASPD for well-measured showers can be considered equal to the projected footprint of the array. Around the threshold however, the collection area is strongly zenith angle and primary energy dependent, with showers fluctuating deep in the atmosphere providing some collection area at the lowest accessible energies. To first approximation the detectability of a $\gamma$-ray shower at ground level depends on the arriving EM energy $E_{\text{em}}$, and the collection area on the fraction of showers that have $E_{\text{em}}$ above a threshold value for a given primary energy and slant depth.
In Figure~\ref{fig:Thres10GeV}, this fraction is shown as a function of slant depth, with an assumed threshold $E_{\text{em}}>$10~GeV.
This threshold has been chosen as an example of an optimistic detection threshold for a future observatory. Figure~\ref{fig:Thres10GeV} compares the results of the full Monte-Carlo shower simulations (markers) with the parameterisations given in Appendix B (lines, see Figures \ref{fig:meanSlantDepthGamma} and \ref{fig:meanSlantDepthProton}). The agreement between the full Monte-Carlo and simple parameterisation is reasonable, illustrating the usefulness of the parameterisations. As for Figures \ref{fig:meanSlantDepthGamma} and \ref{fig:meanSlantDepthProton}, the dependency on slant depth is typically stronger for the $\gamma$-ray induced air showers, which is expected for pure electromagnetic cascades. We would like to remark here that the behaviour is very smooth as a function of slant depth, while individual markers at similar slant depth correspond to a wide range of zenith angles. 

\section{Lateral Extent}

In addition to the total $E_{\text{em}}$ that reaches the ground, the area over which this energy is spread is a crucial parameter in shower detectability for a given array design. We adopt the radius $r_{50}$, in which 50\% of the total electromagnetic energy that reaches the ground is contained, as an indicator of the shower extent.
$r_{50}$ is calculated in the plane perpendicular to the direction of propagation of the primary particle. The choice of 50\% rather than a larger containment fraction is motivated by the large values obtained even for half containment for low energy showers, as described below. We note that the Moli\`ere radius contains approximately 90\% of the energy in the full electromagnetic cascade~\cite{PDGReview}. In air at 5~km above sea level the Moli\`ere radius $R_{m}$ is approximately 130~m.

\begin{figure*}[!ht]
\begin{center}
  \includegraphics[width=0.8\textwidth] {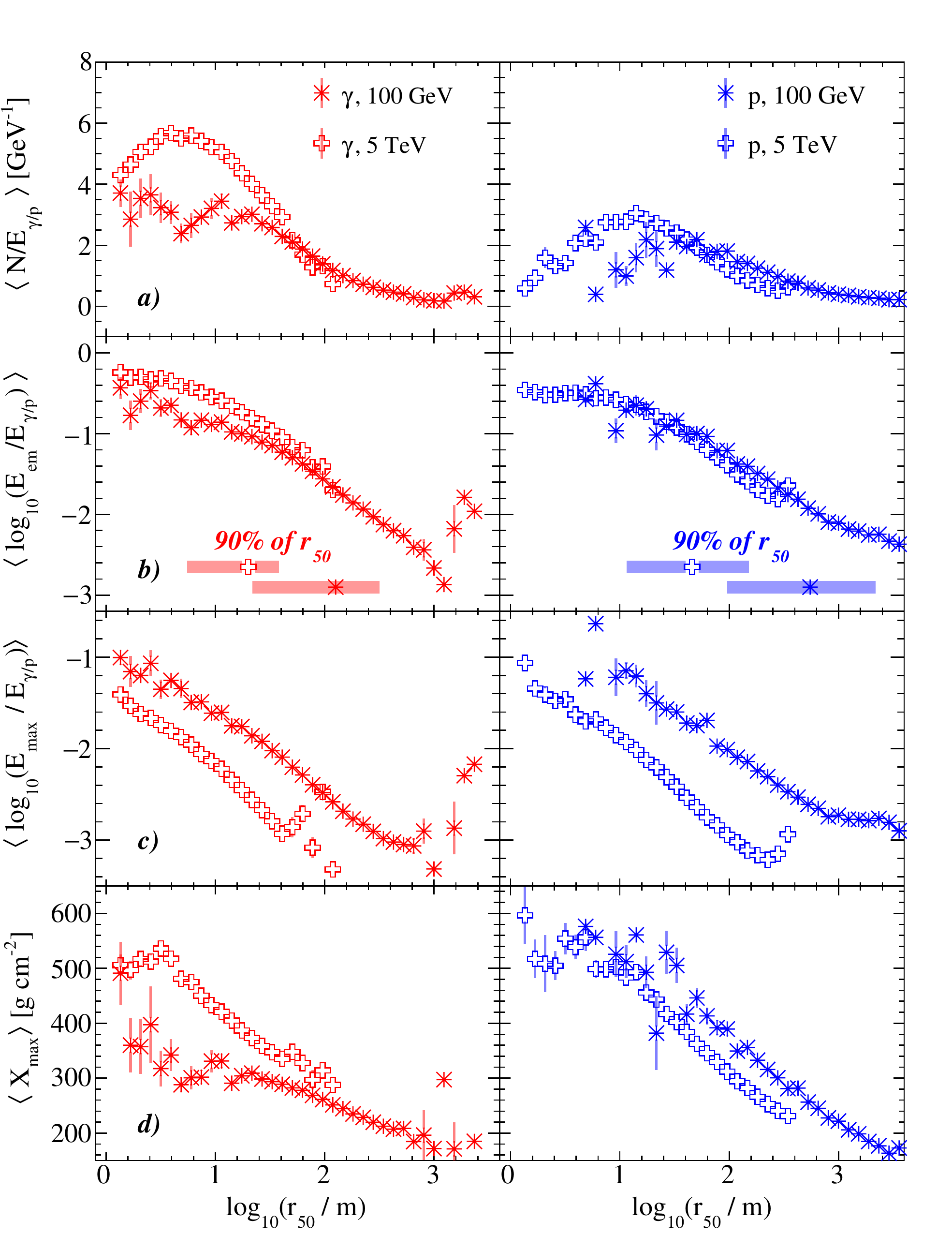}
  \caption{Relationship of the lateral extent parameter $r_{50}$ to other shower parameters for vertical EAS observed at 5~km altitude. Results are shown for $\gamma$-ray (left panels) and proton (right panels) primary particles simulated at two different energies. From top to bottom the following EAS parameters are shown:  \textbf{\textit{a)}} The average number of particles at ground with energy of more than 1~MeV. \textbf{\textit{b)}} Average EM energy at ground. In addition, the bands on this panel indicate the range that contains 90\% of $r_{50}$ values. \textbf{\textit{c)}} The maximum energy of a single EM-particle ($e^\pm,\gamma$). \textbf{\textit{d)}} Depth of shower maximum, obtained from the fit routine within CORSIKA. The parameters shown in \textbf{\textit{a)}}, \textbf{\textit{b)}} and \textbf{\textit{c)}} are scaled with the energy of the primary particle.
  }
\label{fig:r50}
\end{center}
\end{figure*}

Figure~\ref{fig:r50} illustrates the relationship of the parameter $r_{50}$ to several other related EAS characteristics.
For a fixed primary energy, $r_{50}$ depends on the stage of development of the EAS, which can be clearly observed from the correlations with the EAS parameters shown in Figure~\ref{fig:r50}. 
From the behaviour of $r_{50}$ with respect to the number of particles at ground $N$ (panel \textbf{\textit{a}}) it is clear (at least for 5~TeV showers) that the smallest values of $r_{50}$ are provided by showers which reach ground-level before reaching their maximum development\footnote{Note that such showers are rare, lying outside the 90\% range of $r_{50}$ indicated in the inset of Figure~\ref{fig:r50}b.}.
However, this effect is not so clear from the relationship to shower maximum $X_{\text{max}}$ (panel \textbf{\textit{d}}) as obtained by fitting the longitudinal shower development within the simulation package. The reason for this is presumably the difficulty of extrapolating the longitudinal profile to a below-ground $X_{\text{max}}$, in particular for small showers with significant fluctuations.
For showers that reach $X_{\rm max}$ well before arriving to the ground $r_{50}$ does show the expected correlation with shower maximum. 
  
A better handle on the stage of EAS development might be obtained by the energy $E_{\text{max}}$ (panel \textbf{\textit{c}}) of the most energetic single electromagnetic particle in the shower at ground, which correlates to the number of interactions in the EM-cascade before the ground is reached. Whilst $E_{\text{max}}$ is not a straight-forward parameter to measure, Figure~\ref{fig:r50} indicates that both it and 
$r_{50}$ may be very useful observables as indicators of the stage of shower development. This idea is reinforced by the fact that 
the behaviour of $r_{50}$ with the fraction of energy arriving at the ground (Figure \ref{fig:r50} \textbf{\textit{b}}) is very similar for the two primary energies illustrated.


\begin{figure}[!ht]
  \begin{center}
  \includegraphics[width=0.45\textwidth] {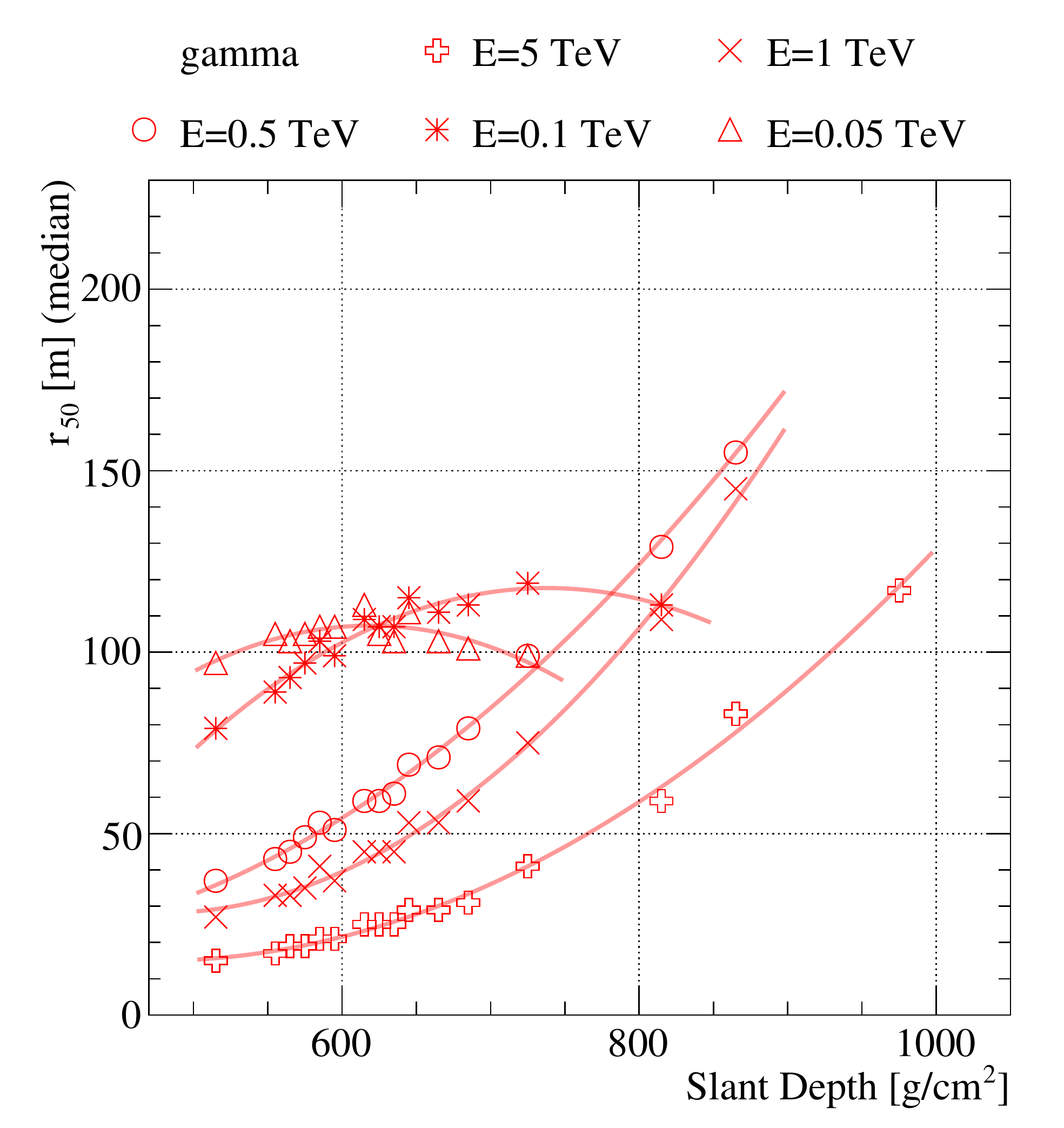}
\includegraphics[width=0.45\textwidth] {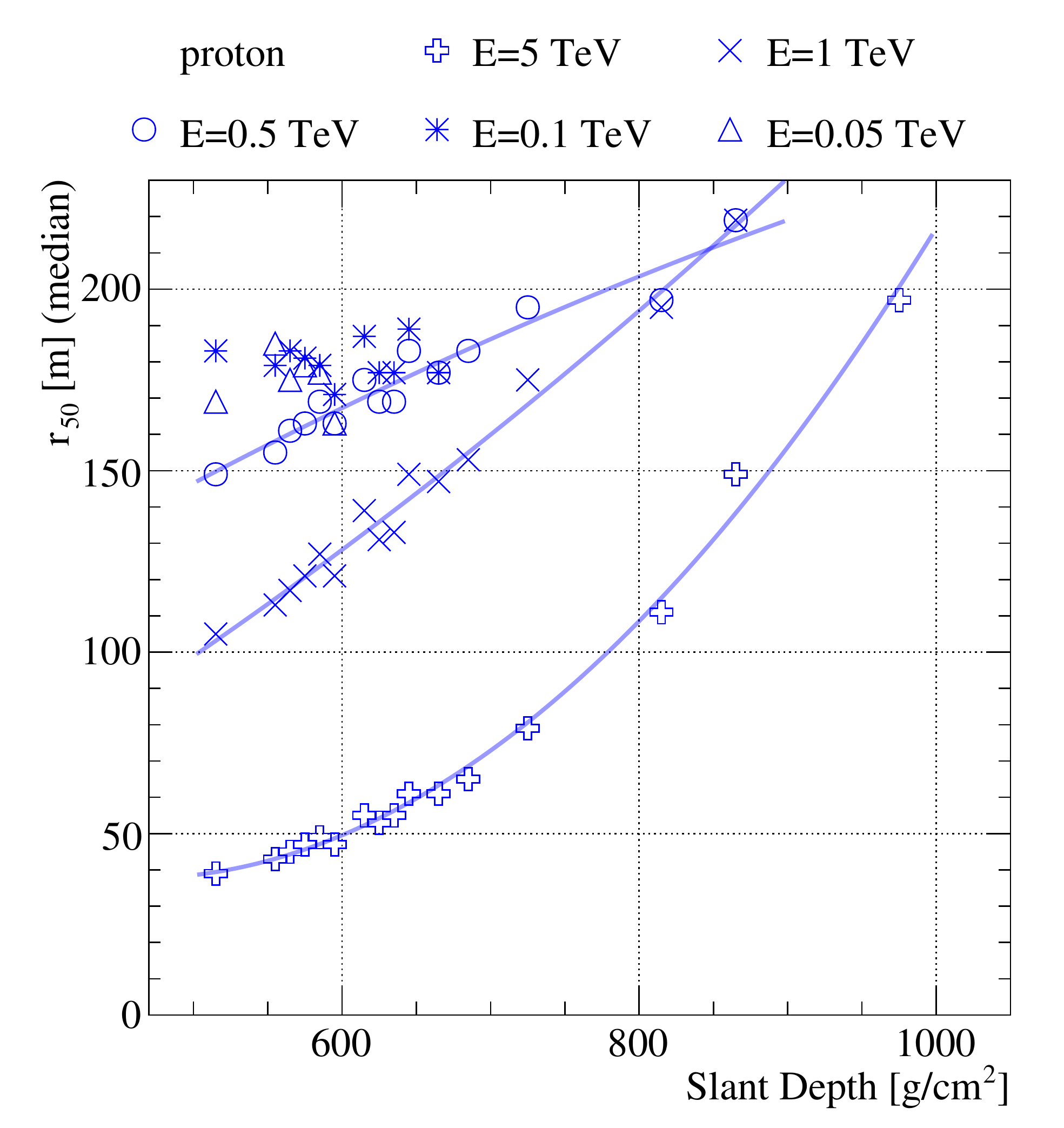}
\caption{Lateral shower extent parameter $r_{50}$ as a function of slant depth. The points correspond to the median value of $r_{50}$. The top panel shows the dependency for $\gamma$-ray showers while the behaviour for protons is shown in the bottom panel. The results of fitting a second order polynomial are shown as solid lines, fit parameters are given in Appendix~B.}
\label{fig:ShowerSize}
\end{center}
\end{figure}

Figure \ref{fig:ShowerSize} shows the dependence of the median
$r_{50}$ on slant depth, indicating again the very large typical
extent of showers well beyond shower maximum. Again the relationship
is fit with a second order polynomial and the best fit parameters
provided in Appendix~B for both gammas and protons. For the lowest
energy showers, fluctuations are very large and the trends are less
clear, but clearly the typical extent is very large in comparison to
the traditional regime of air shower measurements. For higher energy
showers closer to $X_{\rm max}$
we find that the 90\% energy containment radius $r_{90}$ is close to the Moli\`ere radius as expected.

\begin{figure}[!ht]
\begin{center} 
\includegraphics[width=0.5\textwidth] {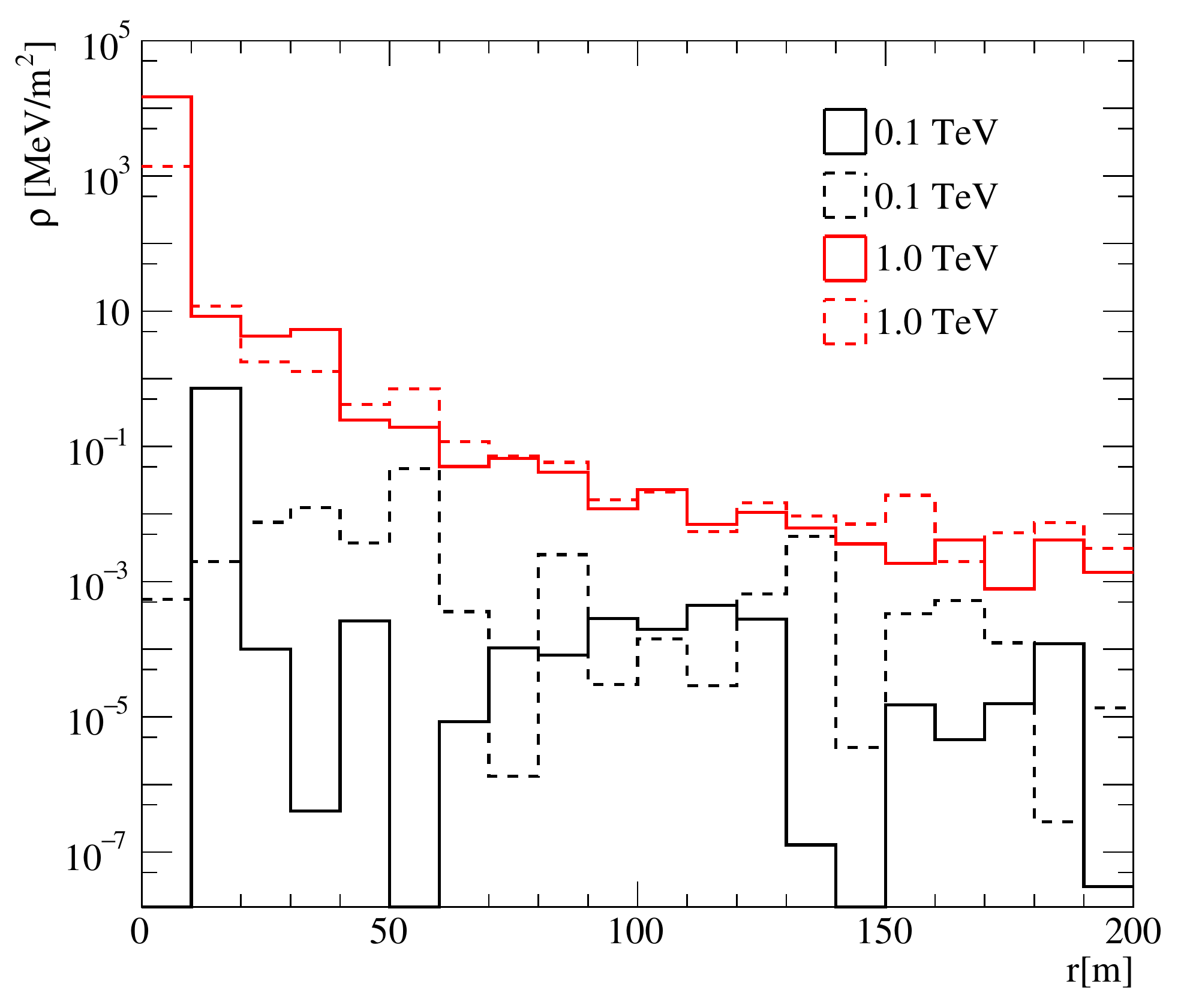}
\caption{Electromagnetic energy density $\rho$ as a function of distance to the shower axis for four sample vertical $\gamma$-ray induced showers of two different energies observed at an altitude of 5~km.}
\label{fig:LDF}
\end{center}
\end{figure}

The deviation from this behaviour as the
shower peters out is consistent with the disappearance of particles at
or above the critical energy from which the Moli\`ere radius is defined,
with subsequent rapid multiple scattering and large displacements from
the shower axis for the remaining low energy electrons. As an example,
considering 100~GeV $\gamma$ rays with a typical $r_{50}$ of 150~m, it is
apparent from Figure~\ref{fig:r50} that such showers have typically
100 particles sharing about 2~GeV of kinetic energy and no particles
with energy $>200$~MeV. Another consequence of the low particle number
and large scattering is that the lateral distribution of particles
becomes very flat and with very large fluctuations. In general we find that $r_{90}\sim 6 r_{50}$, but again with large fluctuations for the low energy cases. To give an indication of the typical appearance of showers of these energies at ground-level, Figure \ref{fig:LDF} compares the lateral distribution of EM
energy for example $\gamma$-ray initiated showers of different
energies. For the 1~TeV examples the distributions are clearly centrally peaked and rather similar, but
at 100~GeV fluctuations are very large, shower-to-shower difference are very large, and no clear shower core may be present.

\section{Muon Number} 

\begin{figure}[ht]
\begin{center}
\includegraphics[width=0.5\textwidth] {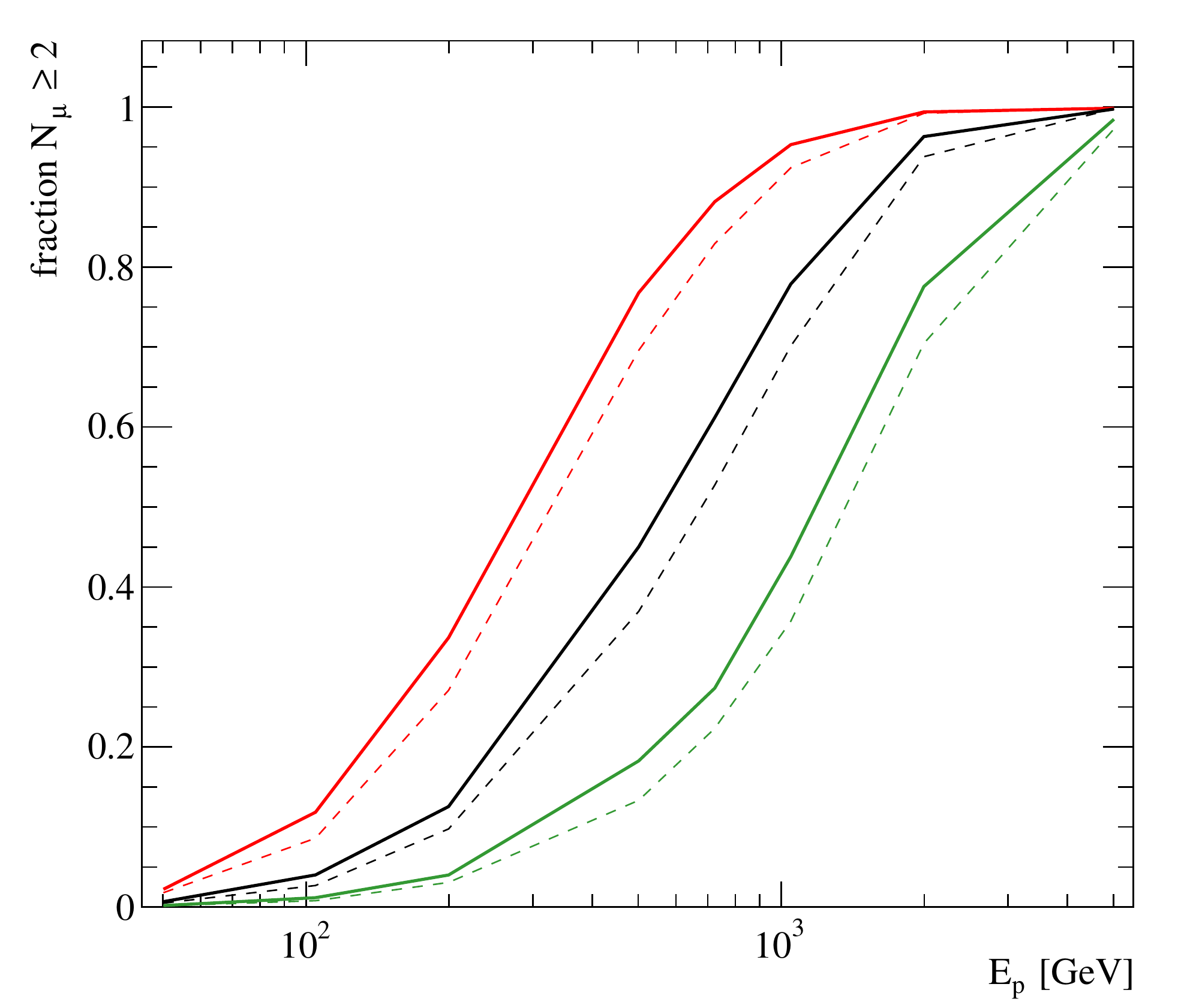}
\includegraphics[width=0.5\textwidth] {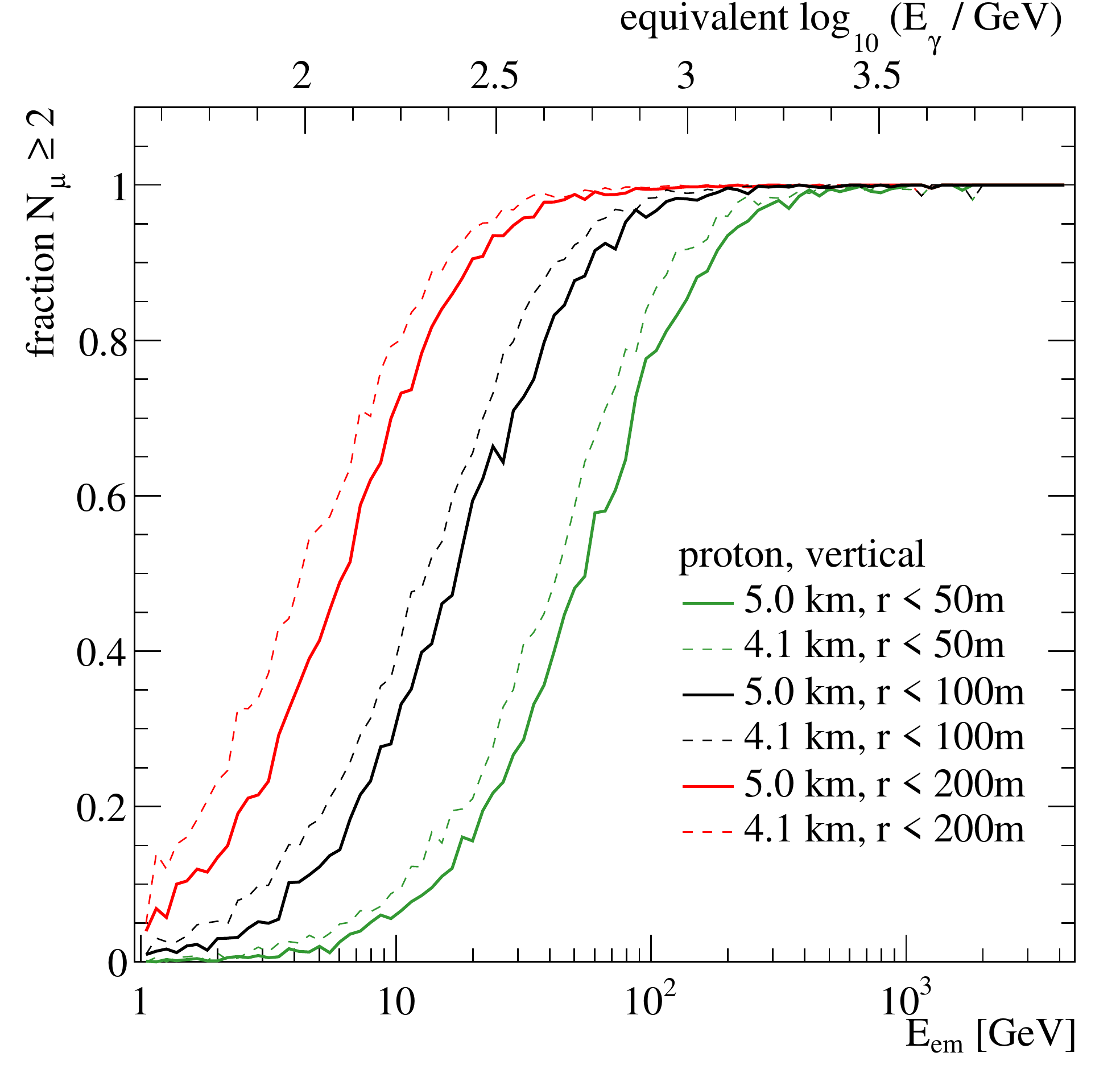}
\caption{Top: fraction of proton induced showers that have more than two muons within 50/100/200~m distance of the shower EM barycentre as a function of primary particle energy. Bottom: the same fraction but as a function of the electromagnetic energy brought to the ground ($E_{\text{em}}$). The upper label for the abscissa indicates the average energy of a vertical incoming $\gamma$-ray primary which results in the same $E_{\text{em}}$ as observed at 5~km altitude.}
\label{fig:fNmu_r}
\end{center}
\end{figure}

As is clear already from Figure~\ref{fig:fNmu_Eem_distr}, the number of muons at ground relative to electromagnetic energy provides a powerful discriminant between $\gamma$-ray and proton showers. As well as the number of arriving muons, the lateral extent of muons is a key consideration. Due to the significant transverse momentum of their parent pions, muons are often present at rather large distances to the shower axis, and indeed for showers in the energy range considered here, the shower axis is often poorly defined based on ground-level observables. Here we consider the distance of muons from the barycentre of the electromagnetic particles in the shower, as the key factor for discrimination is the association of a given muon with the EM-component of the shower. Given the rather high rate of background muons unassociated to any shower at ground, it is unlikely that the association of a single muon can be used for discrimination purposes. In the following we focus on the fraction of showers in which at least two muons reach ground level, within radii, 50~m, 100~m and 200~m, of the EM centroid. Figure~\ref{fig:fNmu_r} illustrates these fractions as a function of both primary energy and electromagnetic energy at ground for proton initiated showers. In addition, we provide here for reference the equivalent $\gamma$-ray energies which result in the same mean $E_{\text{em}}$ for this zenith angle and altitude.

As showers are observed at progressively larger slant depths, the beam of muons broadens and an increasing number of muons decay before reaching the ground, both effects leading to a lower muon count per unit area. The upper panel of Figure~\ref{fig:fNmu_r} illustrates this reduction in muon number per unit area with decreased observation altitude, but as is clear in the lower panel of this Figure, the attenuation effect is much stronger for the electromagnetic component.
It is apparent that a detector linear scale of $\gg$50~m is needed to exploit muons for background rejection for $\gamma$-rays below 1~TeV.

\section{Implications for Array Design}

For altitudes much above 5~km it becomes extremely difficult to find a suitable site for a $\gamma$-ray observatory. It therefore seems likely that a future observatory must be designed to deal with being $\sim$6 radiation lengths below $X_{\rm max}$ even for vertical showers. For primary $\gamma$-rays below 1~TeV this translates to low typical ground-level particle energies and hence as large, and strongly fluctuating, shower footprint~(see Figures~\ref{fig:r50} and \ref{fig:LDF}). Given that the nominal 10~GeV ground-level EM energy threshold assumed for Figure~\ref{fig:Thres10GeV} is already extremely challenging, an array on a scale significantly larger than $r_{50}$ (i.e. $\sim$100~m), and with fill factor approaching 100\%, is needed for $\gamma$-ray astronomy at these energies. For hadron rejection based on muon identification the relevant scale is even larger as can be seen from Figure~\ref{fig:fNmu_r}.
A detector on the scale of HAWC detector (array area $\sim$20,000~m$^{2}$) will typically detect only a modest fraction of the arriving EM energy of a sub-TeV shower and a small fraction of the muons in a background TeV proton shower. Given the absence of a clear core for many far-beyond $X_{\rm max}$ low energy showers (see e.g.~Figure~\ref{fig:LDF})
morphology-based hadron rejection will be extremely challenging and clear muon identification appears to be a promising approach for most of the energy range considered here. Figure~\ref{fig:fNmu_r} indicates that muon counting can be exploited for $\gamma$-ray energies as low as 300~GeV with a sufficiently large array.

\section{Conclusions}

The parameterisations of proton and $\gamma$-ray showers at ground level presented here should prove useful in the design of next generation $\gamma$-ray observatories based on ground-particle detection. In the shower tail, far beyond $X_{\rm max}$ where sub-TeV measurements are possible, the typical extent of showers becomes very large and large array footprint as well as close to 100\% fill factor is required. The tagging of individual muons offers an opportunity for background rejection even in the sub-TeV regime, where the very large fluctuations in $\gamma$-ray showers at ground will make rejection based on the lateral distribution of particles very difficult.


\bibliographystyle{spphys_ruben} 
\bibliography{./references}

\begin{thebibliography}{10}
\providecommand{\url}[1]{{#1}}
\providecommand{\urlprefix}{URL }
\expandafter\ifx\csname urlstyle\endcsname\relax
  \providecommand{\doi}[1]{DOI \discretionary{}{}{}#1}\else
  \providecommand{\doi}{DOI \discretionary{}{}{}\begingroup
  \urlstyle{rm}\Url}\fi

\bibitem{HAWC_CRAB}
Abeysekara, A.~U., et~al., The Astrophysical Journal,  \textbf{843}(1), 39
  (2017).
\newblock \urlprefix\url{http://stacks.iop.org/0004-637X/843/i=1/a=39}

\bibitem{HESS_Performance}
{Aharonian}, F., et~al., \aap,  \textbf{457}, 899 (2006).
\newblock \doi{10.1051/0004-6361:20065351}

\bibitem{MAGIC_Performance}
{Aleksi{\'c}}, J., et~al., Astroparticle Physics,  \textbf{72}, 76 (2016).
\newblock \doi{10.1016/j.astropartphys.2015.02.005}

\bibitem{impact}
{Parsons}, R.~D. et~al., Astroparticle Physics,  \textbf{56}, 26  (2014).
\newblock \doi{10.1016/j.astropartphys.2014.03.002}

\bibitem{HAWC_Energy}
{Marinelli}, S. et~al., in \emph{35th International Cosmic Ray Conference,
  PoS(ICRC2017)714} (2017).
\newblock \urlprefix\url{https://pos.sissa.it/301/714/}

\bibitem{engel_air_showers}
{Engel}, R., et~al., Annual Review of Nuclear and Particle Science,
  \textbf{61}(1), 467 (2011)

\bibitem{Tibet_Survey}
{Amenomori}, M., et~al., \apj,  \textbf{633}, 1005 (2005).
\newblock \doi{10.1086/491612}

\bibitem{ARGO_Performance_2006}
{Aielli}, G. et~al., Nuclear Instruments and Methods in Physics Research A,
  \textbf{562}, 92 (2006)

\bibitem{Milagro_Crab}
{Abdo}, A.~A., et~al., ApJ,  \textbf{750}, 63 (2012).
\newblock \doi{10.1088/0004-637X/750/1/63}

\bibitem{HAWC_Performance_2013}
{Abeysekara}, A.~U., et~al., Astroparticle Physics,  \textbf{50}, 26 (2013).
\newblock \doi{10.1016/j.astropartphys.2013.08.002}

\bibitem{HAWC_survey}
Abeysekara, A.~U., et~al., The Astrophysical Journal,  \textbf{843}(1), 40
  (2017).
\newblock \urlprefix\url{http://stacks.iop.org/0004-637X/843/i=1/a=40}

\bibitem{LATTES_ICRC}
{Concei\c{c}\~ao}, R. et~al., in \emph{35th International Cosmic Ray
  Conference, PoS(ICRC2017)784} (2017).
\newblock \urlprefix\url{https://pos.sissa.it/301/784/}

\bibitem{ALTO_ICRC}
{Becherini}, Y., et~al., in \emph{35th International Cosmic Ray Conference,
  PoS(ICRC2017)782} (2017).
\newblock \urlprefix\url{https://pos.sissa.it/301/782/}

\bibitem{ALPACA_ICRC}
Ohnishi, M., in \emph{35th International Cosmic Ray Conference,
  PoS(ICRC2017)827} (2017).
\newblock \urlprefix\url{https://pos.sissa.it/301/827/}

\bibitem{ARGO_ICRC}
Di~Sciascio, G., et~al., in \emph{35th International Cosmic Ray Conference,
  PoS(ICRC2017)782} (2017).
\newblock \urlprefix\url{https://pos.sissa.it/301/782/}

\bibitem{HARM_ICRC}
Schoorlemmer, H., et~al., in \emph{35th International Cosmic Ray Conference,
  PoS(ICRC2017)819} (2017).
\newblock \urlprefix\url{https://pos.sissa.it/301/819/}

\bibitem{SCIENCECASE_ICRC}
Mostafa, M., et~al., in \emph{35th International Cosmic Ray Conference,
  PoS(ICRC2017)851} (2017).
\newblock \urlprefix\url{https://pos.sissa.it/301/851/}

\bibitem{corsika}
{Heck}, D., et~al., \emph{{CORSIKA: a Monte Carlo code to simulate extensive
  air showers.}} (Forschungszentrum Karlsruhe GmbH, 1998)

\bibitem{Heitler_model}
{Heitler}, W., \emph{{Quantum theory of radiation}} (International Series of
  Monographs on Physics, Oxford: Clarendon, 3rd ed., 1954)

\bibitem{Rossi_model}
{Rossi}, B., \emph{{High Energy Particles}} (Prentice-Hall, Englewood Cliffs,
  NJ, 1952)

\bibitem{PDGReview}
Tanabashi, M., et~al., Phys. Rev. D,  \textbf{98}, 030001 (2018).
\newblock \doi{10.1103/PhysRevD.98.030001}

\end{thebibliography}

\section*{Appendix A}
\begin{figure}[!ht]
\begin{center} %
\includegraphics[width=0.5\textwidth]{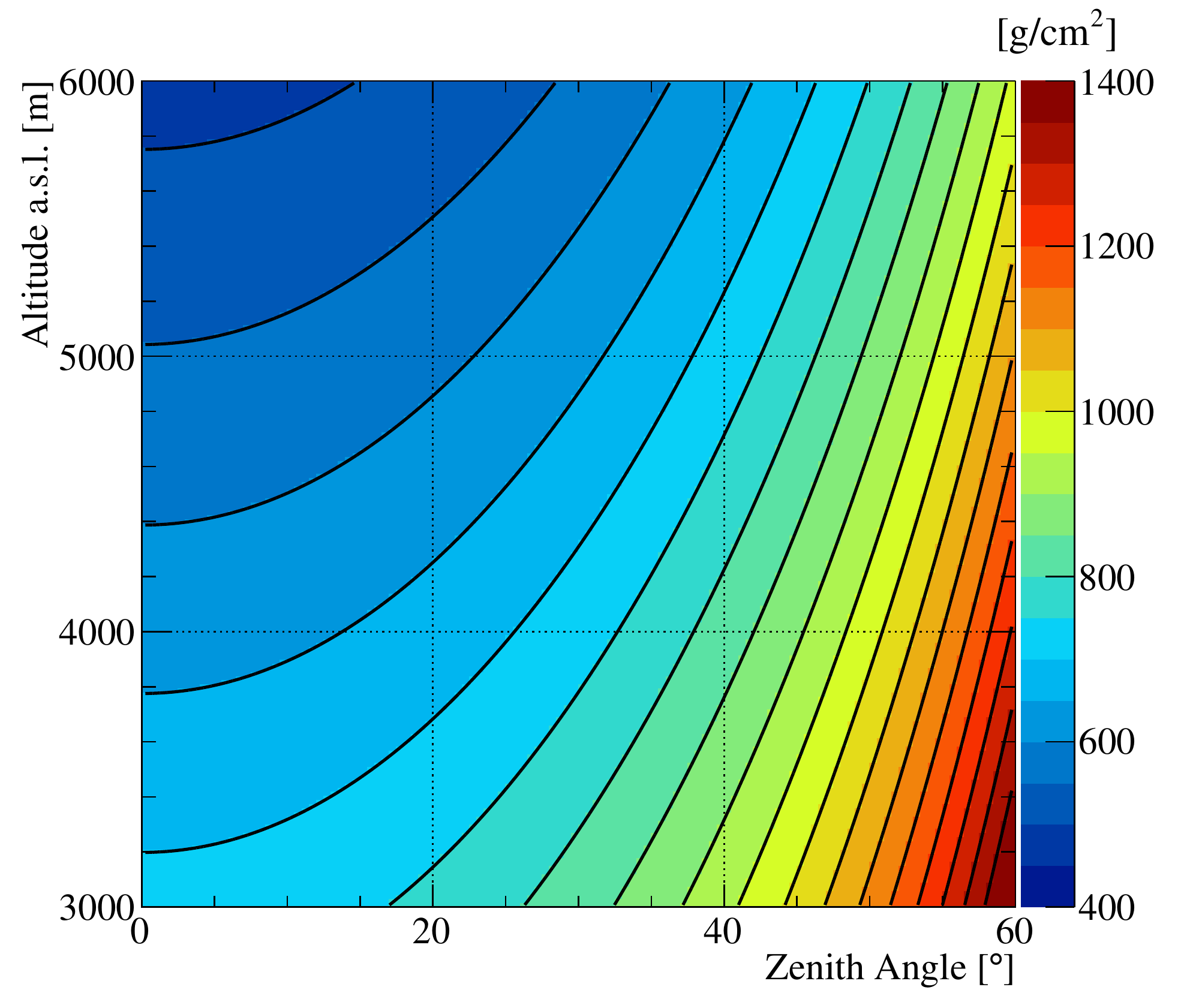} 
\caption{Slant depth at ground level as a function of altitude and zenith angle for a US standard atmosphere. Each contour indicates a 50~g~cm$^{-2}$ change.}
\label{fig:slant_depth}
\end{center}
\end{figure}
Figure~\ref{fig:slant_depth} provides a reference for slant depth $X(z,\theta) = \int_\infty^{z}\rho_{\mathrm{atm}}(z',\theta)\cos\theta dz'$ versus zenith angle and observation altitude for a US standard atmosphere.

\section*{Appendix B}
In Table \ref{tab:fitEem} we include the fit parameters from Figures \ref{fig:meanSlantDepthGamma}  and \ref{fig:meanSlantDepthProton}. These parameters can be used to derive the mean and standard distribution of the distributions at a given energy for gamma and proton primaries and the selected slant depth. 
\begin{table}[!ht]
\begin{tabular}{ c|c|c|c }
$\langle \log_{10}\left(\frac{E_{\text{em}} }{E_{\gamma}} \right) \rangle$ & $p_{0}$&$p_{1}$&$p_{2}$\\
\hline
50 GeV &6.18e-01&	-2.51e-03&	-3.20e-06\\
 100 GeV &  5.32e-01&	-1.99e-03&	-3.11e-06\\
 500 GeV &  5.46e-01&	-1.49e-03&	-2.70e-06\\
 1000 GeV & 5.47e-01&	-1.30e-03&	-2.57e-06\\
 5000 GeV & 3.63e-01&	-3.92e-04&	-2.73e-06\\
\hline
$\sigma(\log_{10}\left(\frac{E_{\text{em}} }{E_{\gamma}} \right))$ & $p_{0}$&$p_{1}$&$p_{2}$\\
\hline
50 GeV &  2.79e-01&	-6.29e-04&	1.68e-06\\
 100 GeV &  3.38e-01&	-6.84e-04&	1.36e-06\\
 500 GeV & 7.29e-02&	1.84e-04&	3.58e-07\\
 1000 GeV &-1.05e-01&	6.84e-04&	-7.21e-08\\
 5000 GeV & -1.65e-01&	7.33e-04&	-1.61e-07\\
 \hline
 $\langle \log_{10}\left(\frac{E_{\text{em}} }{E_{p}} \right) \rangle$ & $p_{0}$&$p_{1}$&$p_{2}$\\
 \hline
50 GeV&   -2.93e-02&	-3.90e-03&	-\\
 100 GeV& 2.51e-01&	-4.01e-03&	-\\
 500 GeV &  7.14e-01&	-4.47e-03&	6.92e-07\\
 1000 GeV &6.49e-01&	-3.84e-03&	2.87e-07\\
 5000 GeV &2.57e-01&	-1.73e-03&	-1.11e-06\\
\hline
$\sigma(\log_{10}\left(\frac{E_{\text{em}} }{E_{p}} \right))$ & $p_{0}$&$p_{1}$&$p_{2}$\\
\hline
50 GeV &2.19e-01&	9.64e-04&	-\\
 100 GeV&-1.14e-02&	1.19e-03&	-\\
 500 GeV&1.55e-01&	4.63e-04&	2.36e-07\\
 1000 GeV & -4.82e-03&	9.17e-04&	-1.98e-07\\
 5000 GeV &-3.45e-01&	1.63e-03&	-7.26e-07\\
\end{tabular}
\caption{Fit results to parameterise the distributions of $E_{\text{em}}$ as a function slant depth. Listed are the parameters from a polynomial fit f$(X_{\text{sl}}) = p_0 + p_1 X_{\text{sl}} + p_2 X_{\text{sl}}^2$, where $X_{\text{sl}}$ is the slant depth, and $f$ is given in the first column of the header of each sub-table.}
\label{tab:fitEem}
\end{table}

In Tables \ref{tab:r50G} and \ref{tab:r50P} we include the fit parameters from Figure \ref{fig:ShowerSize}. These parameters can be used to estimate shower sizes as a function of slant depth.  Especially for low energy results, it is not wise to extrapolate these functions outside the range indicated in Figure \ref{fig:ShowerSize}.
\begin{table}[!ht]
\begin{tabular}{ c|c|c|c }
 $E_{\gamma}$& $p_{0}$&$p_{1}$&$p_{2}$\\
\hline
 50 GeV: &-2.36e+02&	1.11e+00&	-8.97e-04\\
 100 GeV: &  -3.11e+02&	1.16e+00&	-7.87e-04\\
 500 GeV: &6.68e+01&	-2.98e-01&	4.63e-04\\
 1000 GeV: &  1.99e+02&	-7.18e-01&	7.53e-04\\
 5000 GeV: &1.06e+02&	-3.86e-01&	4.09e-04\\
\end{tabular}
\caption{Best fit parameters for the function $r_{50}(X_{\text{sl}}) = p_0 + p_1 X_{\text{sl}} + p_2 X_{\text{sl}}^2$ for $\gamma$-ray initiated air showers. $X_{\text{sl}}$ is the slant depth.}
\label{tab:r50G}
\end{table}
\begin{table}[!ht]
\begin{tabular}{ c|c|c|c }
 $E_{p}$& $p_{0}$&$p_{1}$&$p_{2}$\\
\hline
50 GeV: &1.75e+01&	-&	-\\
 100 GeV: &1.75e+01&	-&	-\\
 500 GeV: &1.75e+01&	3.01e-01&	-8.54e-05\\
 1000 GeV: &  -1.14e+01&	1.60e-01&	1.21e-04\\
 5000 GeV: &1.68e+02&	-5.68e-01&	6.16e-04\\
\end{tabular}
\caption{Best fit parameters for the function $r_{50}(X_{\text{sl}}) = p_0 + p_1 X_{\text{sl}} + p_2 X_{\text{sl}}^2$ for proton initiated air showers. $X_{\text{sl}}$ is the slant depth.}
\label{tab:r50P}
\end{table}

\end{document}